# Determination of SNe explosions frequency distribution function.

## Method and numerical simulations

A.A.Akopian

Byurakan Astrophysical Observatory, Byurakan, 378433, Armenia
email : aakopian@bao.sci.am

**ABSTRACT**

*Aims.* The method for determination of the Supernovae (SNe) explosions frequency distribution function based on the assumption of explosions independence are offered.

*Methods.* The method is based on assumption that the sequence of SNe explosions in an individual galaxy is a Poisson sequence. The essence of the method is in the determination of statistical moments of the frequency of the SNe explosions and subsequent determination of distribution function .

*Results.* The program of numerical simulation has been developed for testing the efficiency of the method. Numerical simulations show that even for a small mean number of registered SNe explosions, method allows restoring initial distribution function. The results of numerical simulations are given.

## 1. Introduction

During an explosion of a supernova star (SN) huge amount of energy is released. Therefore it is not surprising, that SNe play the important roles in many astrophysical theories and models. However despite of the big attention to the SN, many of key problems, of both theoretical and observational characters, still are far from the decision. The some features of SNe (unpredictability and a rarity of events, greater distances, etc.) seriously restrict collecting and processing of statistically homogeneous and rich materials.

On the other hand, as a result of the SNe surveys carried out in the past years or still ongoing, significant amount of data on SNe in other galaxies has gathered. Systematic surveys allow to compute the SNe mean rates in various galaxies samples, by applying the control time technique (Zwicky 1942), after taking into account the various selection effects (see for example , Cappellaro et al. 1999, Filippenko et al. 2001). Usually the SNe rate is normalized to some parameter of the galaxies related to their "sizes". The most common quantities used are the luminosities in different (optical or IR) bands.

The mean rate of SNe is an important characteristic of the galaxies sample, however it is obviously that the most productively could be determination of SNe rate distribution function. Direct determination of this function on the basis of SNe registration now is impossible because of small number $(1-2)$ of SNe explosion observed in separate galaxies. For more or less confident determination it would be necessary to register at least $\sim 10$ SNe in each galaxy. Because of the low mean rate $(< 10^{-2} y^{-1})$ of SNe stars one can conclude that direct determination of distribution function is impossible during the next few centuries. In this study we present the method which allows to determine required function using lesser quantity of data. The main idea of the method is the determination of statistical moments of the SNe explosions frequency (hereinafter instead of term "SNe rate" we shall use the term "SN explosions frequency", which is more convenient in our approach presented below)and subsequent determination of distribution function.

In section 2 we write down the expressions for statistical moments of SNe explosions frequency. For the determination of the statistical moments of distribution function the randomness and independence of SNe explosion in the given galaxy is assumed. Under these assumptions the SNe events may be described by a Poisson distribution (see for example Ambartsumian 1968, Richmond, Filippenko, Galisky 1998). Such assumption was made by Ambartsumian for finding flare stars flares frequency distribution function. Ambartsumian had assumed that similar methods can be developed for SNe also (Ambartsumian 1968). Using this approach Akopian (Akopian 1996) had suggested the method for determination of SNe explosion frequency distribution function . In the method the statistics of the time instants of the first SNe explosions was used. The sought distribution function of SNe explosions frequency turned out as inverse-Laplace transformation of the survival function of the time instant of first SNe explosion .

The method offered by Akopian requires essentially smaller quantity of data, but has not applied, because of very specific characters of data needed. In the present work we suggest a new method of determination of SNe explosions frequency distribution function.

## 2. Statistical moments of SNe explosions frequency

Some astrophysical objects manifest irregular, eruptive brightness variations, whose time behavior is well described by a stationary Poisson process. Among these objects are flare stars, and just as galaxies, if considering them as "flaring objects" whose flares are SNe explosions within a given galaxy (Ambartsumian 1988).

Let us suppose that the sequence of SNe explosions in an individual galaxy are a Poisson sequence, i.e., the probability of $k$ SNe explosions over an observation time (total control time) $t$ is:

$$p_k = \frac{(\nu t)^k e^{-\nu t}}{k!} \quad (1)$$

where $\nu$ is SNe explosions frequency (rate). If sample's galaxies has not the same SNe explosions frequency, then:

$$p_k = \int \varphi(\nu) \frac{(\nu t)^k e^{-\nu t}}{k!} d\nu \quad (2)$$

where $\varphi(\nu)$ is the searched distribution function density.

Hence, the theoretical statistical moments of $k$ have the following forms:

$$\mu k_1 = \sum_{k=0}^{\infty} k p_k = \sum_{k=0}^{\infty} k \int \varphi(v) \frac{(vt)^k e^{-vt}}{k!} dv \tag{3}$$

$$\mu k_j = \sum_{k=0}^{\infty} (k - \mu k_1)^j p_k \tag{4}$$

or

$$\mu k_j = \sum_{k=0}^{\infty} (k - \mu k_1)^j \int \varphi(v) \frac{(vt)^k e^{-vt}}{k!} dv, \quad j = 2, 3... \tag{5}$$

Relations between statistical moments of SNe events number and SNe explosions frequency (rate) follows from Eq. (1-5). Changing the order of integration and summation, we can write down:

$$\mu v_1 = \frac{\mu k_1}{t} \tag{6}$$

$$\mu v_2 = \frac{\mu k_2 - \mu k_1}{t^2} \tag{7}$$

$$\mu v_3 = \frac{\mu k_3 - 3\mu k_2 + 2\mu k_1}{t^3} \tag{8}$$

$$\mu v_4 = \frac{\mu k_4 - 6\mu k_3 - 6\mu k_3 \cdot \mu k_1 + 11\mu k_2 - 6\mu k_1 + 3\mu k_1^2}{t^4} \tag{9}$$

Here

$$\mu v_1 \equiv \int v \varphi(v) dv \tag{10}$$

$$\mu v_j \equiv \int (v - \mu v_1)^j \varphi(v) dv, \quad j = 2, 3, 4 \tag{11}$$

are statistical moments of explosions frequency, particulary $\mu v_1$ is equivalent to mean rate. Accordingly $\mu v_2, \mu v_3, \mu v_4$ are the higher central moments. The empirical statistical moments of SNe number can be derived from observational data:

$$\widehat{\mu k_1} = \sum_{k=0}^{\infty} k \widehat{p_k} \tag{12}$$

$$\widehat{\mu k_j} = \sum_{k=0}^{\infty} (k - \widehat{\mu k_1})^j \widehat{p_k}, \quad j = 2, 3, 4 \tag{13}$$

where $\widehat{p_k}$ is the part of galaxies with $k$ SNe. Having substituted the empirical moments (12,13) instead of theoretical (6-9), moments of SNe explosion frequency are determined.

## 3. SNe Explosions frequency distribution function

For SNe explosions distribution function determination we involve Pearson's distributions. The densities of the Pearson's families distributions obey the equation (Kendall, Stuart 1966):

$$\frac{d\varphi}{dv} = \frac{(v - a)\varphi}{b_0 + b_1 v + b_2 v^2}$$

The coefficients $a, b_i$ can be expressed via the first four moments of the distribution by the following way:

$$a = \frac{-\mu v_3(\mu v_4 + 3\mu v_2^2)}{10\mu v_2 \mu v_4 - 18\mu v_2^3 - 12\mu v_3^2}$$

**Table 1.** The D, $\lambda$ values for common distributions

| Distribution | D | $\lambda$ |
|---|---|---|
| Gamma | < 0 | $\infty$ |
| Beta | < 0 | < 0 |
| Exponential | 0 | 0 |
| Normal | 0 | - |
| Pearson's IV | > 0 | (0, 1) |

**Table 2.** The parameters of common distributions

| Distribution density | View | Parameters $(p_1, p_2, p_3, p_4)$ |
|---|---|---|
| Gamma | $p_1^{p_2} (\nu - p_3)^{p_2-1} \frac{e^{p_1(-\nu+p_3)}}{\Gamma(p_2)}$ | $\frac{2\mu\nu_2}{\mu\nu_3}, \frac{4\mu\nu_2^3}{\mu\nu_3^2}, \mu\nu_1 - \frac{2\mu\nu_2^2}{\mu\nu_3}$ |
| Beta | $\frac{\Gamma(p_1+p_2+2)}{\Gamma(p_1+1)\Gamma(p_2+2)} \left[ \frac{(p_3+\nu)^{p_1}(p_4-\nu)^{p_2}}{(p_3+p_4)^{p_1+p_2+1}} \right]$ | —[a] |
| Exponential | $p_1 e^{p_1(-\nu+p_2)}$ | $\frac{2\mu\nu_2}{\mu\nu_3}, \mu\nu_1 - 1$ |
| Normal | $\frac{1}{\sqrt{2\pi}p_2} e^{-\left(\frac{\nu-p_1}{p_2}\right)^2}$ | $\mu\nu_1, \sqrt{\mu\nu_2}$ |

[a] Are derived as numerical solution of system of 4 equations

$$b_0 = \frac{\mu\nu_2(4\mu\nu_2\mu\nu_4 - 3\mu\nu_3^2)}{10\mu\nu_2\mu\nu_4 - 18\mu\nu_2^3 - 12\mu\nu_3^2}$$

$$b_1 = \frac{-\mu\nu_3(\mu\nu_4 + 3\mu\nu_2^2)}{2(10\mu\nu_2\mu\nu_4 - 18\mu\nu_2^3 - 12\mu\nu_3^2)}$$

$$b_2 = \frac{2\mu\nu_2\mu\nu_4 - 3\mu\nu_3^2 - 6\mu\nu_2^3}{10\mu\nu_2\mu\nu_4 - 18\mu\nu_2^3 - 12\mu\nu_3^2}$$

Type of searched distribution is determined by values:

$$D = b_0 b_2 - b_1^2$$

$$\lambda = \frac{b_1^2}{b_0 b_2}$$

The $D, \lambda$ values for common distribution are given in Table 1. As a whole 12 type of Pearson's distribution are distinguished. Their parameters are expressed via statistical moments. The parameters of some common distributions are given in Table 2.

Thus, determination of sought function involves the following steps:

– Determination of the first four empirical moments of SN number (12-13).
– Computing of the first four moments of SN explosions frequency (6-9).
– Computing of values of $D$ and $\lambda$ and thereby determination of the distribution type.
– Equating of the frequency's empirical and theoretical moments.
– Solution of the resulting equations for the unknown parameters and, final determination of the unknown distribution.

## 4. Numerical simulations

The mathematical program of numerical simulation was developed in order to test the efficiency of a method. The program was created by means of Mathcad software. Random values of SNe explosions frequency $v_i$, ($i = 1, 2...n$, where $n$ is number of galaxies in sample) obeying some distribution function $\varphi(v)$, are generated using random number generators. Then, Poisson random number generator generates random number of SN explosion $k_i$ for each galaxy at frequency $v_i$ over random total control time $t_i$. Applying the method described above, for the given numbers $k_i$, one can determine empirical distribution function $\widehat{\varphi(v)}$, and compare it with initial one $\varphi(v)$.

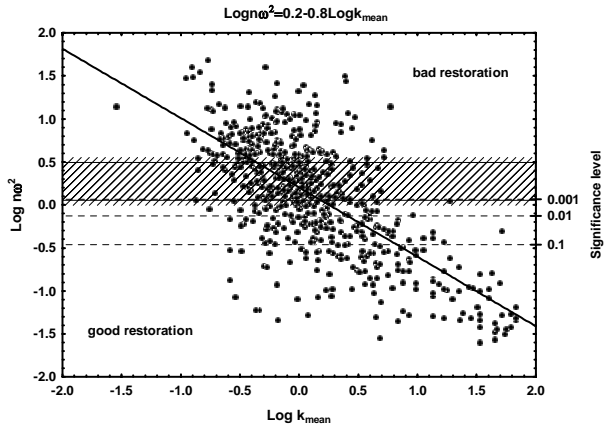

**Fig. 1.**

One must note that each galaxy has a own set of total control time (one for every type of SN light curve), which differs from others. There are two ways to take account inequality of total control times. First way is cutoff of all total control times up to minimal common value, but in this case significant amount of information will be lost. Other way is to take instead of $k_i$ the number of SNe explosions per unit time, i.e. $k_i/t_i$. It seems mathematically incorrect, but numerical simulations show that in practice such procedure have not substantial influence on the function restoration quality.

For comparison of initial and restored functions $\omega^2$ test is used:

$$\omega^2 = \int_{-\infty}^{\infty} (F_{i,n}(x) - F(x))^2 dF(x)$$

where $F_{i,n}$ is an initial function of distribution, and $F$-restored. The mean and a dispersion of statistics $\omega^2$ do not depend on kind $F$ and accept values:

$$M(\omega^2) = \frac{1}{6n} \qquad (14)$$

$$D(\omega^2) = \frac{4n-3}{180n^3} \qquad (15)$$

where $n$ - size of sample (in this case number of galaxies in sample).

Actually distribution $\omega^2$ as a whole does not depend on a kind of distribution function. Statistics $\omega^2$ can be presented in the form of (Kendall, Stuart 1967):

$$n\omega^2 = \frac{1}{12n} + \sum_{r=1}^{n} \left\{ F(x_{(r)}) - \frac{2r-1}{2n} \right\}^2 \qquad (16)$$

where $x_{(r)}$ - ordered statistics $x_{(1)} \leq x_{(2)} \leq ......... \leq x_{(n)}$. Apparently from (14-16) the statistics $n\omega^2$ at even small values of $n$ practically does not depend from $n$. At a preliminary stage of researches other test criteria have been used also, however eventually because of the set forth above properties, this criterion has been chosen.

**Table 3.** Percentages of good , satisfactory and bad restorations

| $lgk_{mean}$ | good | satisfactory | bad |
|---|---|---|---|
| 1.5 | 98.5 | 0.5 | 1 |
| 1.0 | 91 | 7.5 | 1.5 |
| 0.5 | 70 | 22 | 22 |
| 0.0 | 40 | 33 | 27 |
| -0.5 | 14 | 42 | 44 |
| -1.0 | 3 | 16 | 81 |

On Fig. 1 the results of one numerical experiment are shown. During this experiment for $n = 200, 600, 1200$ simulations (by 200 for each) have been carried out. On abscissa the values of the logarithm of a mean "SNe" number in the given sample are brought , on the left ordinate - values of the logarithm $n\omega^2$, on right ordinate - corresponding significance values (only some often used) . Apparently linear dependence takes place:

$$lgn\omega^2 = 0.2 - 0.8 \lg k_{mean} \tag{17}$$

From the data analysis follows, that for the fixed value $k_{mean}$ $lgn\omega^2$ has normal distribution with mean value $m = 0.2 - 0.8 lgk_{mean}$ and standard deviation $\sigma = 0.5$:

$$\psi\left(\lg n\omega^2\right) = \frac{1}{\sqrt{2\pi}\sigma} \exp -\left\{\frac{\lg n\omega^2 - \mu}{\sigma}\right\}^2 \tag{18}$$

Fig.1 is divided into three areas: area (the significance value is $> 0.001$) of good restoration of initial function, area of bad restoration ($n\omega^2 > 3$) and the shaded intermediate area , where from the point of view of mathematical statistics is bad restoration, but at this stage of researches one can consider such level of restoration qualitatively successful. Examples of restoration from each area at $n = 600$ are shown on Fig(2-4) . On these figures the initial and restored distribution functions are presented, and corresponding density functions and "SN" number distribution are depicted also .

The bin's number was calculated according to Sturges's rule (Sturges 1926). The frequency unit plotted on abscisses in the certain sense is arbitrary. It can be considered as SNe events per unit time (for example, in $10^3 years$) or as the same number normalized to luminosity, etc. Using relations (17-18), we have computed percentages of good, satisfactory and bad restorations. These percentages are presented in the Table 3.

Results of the presented numerical experiment are typical. The similar numerous experiments, carried out under other conditions (different types of initial functions, different sets of observational time, different sizes of samples) have shown nearly the same results. The general conclusion is when mean SNe number in a galaxies sample is $\sim 0.5 \div 1$, then method allows to restore initial function more than in 50% cases. Method is not applicable for lower values of mean number.

## 5. Conclusions

1. The method for determination of the Supernovae (SNe) explosions frequency distribution function based on the assumption of explosions independence are suggested.The method is based on assumption that the sequence of SNe explosions in an individual galaxy is a Poisson sequence.

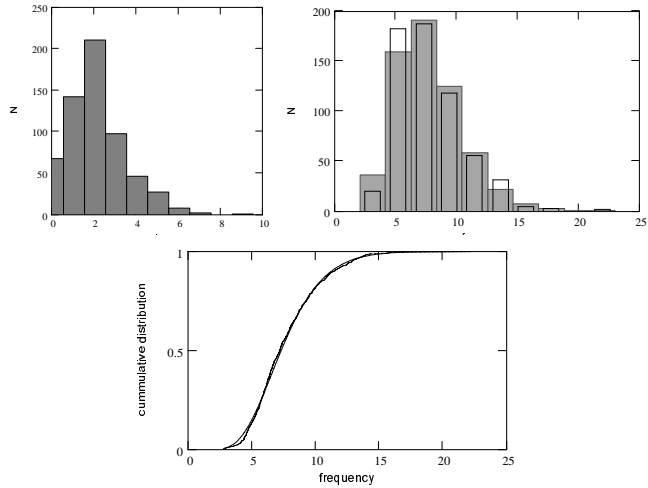

**Fig. 2.** Good restoration. Generated SN number distribution ; Initial (bar) and restored(solid bar) density functions; Initial (broken line) and restored distribution functions

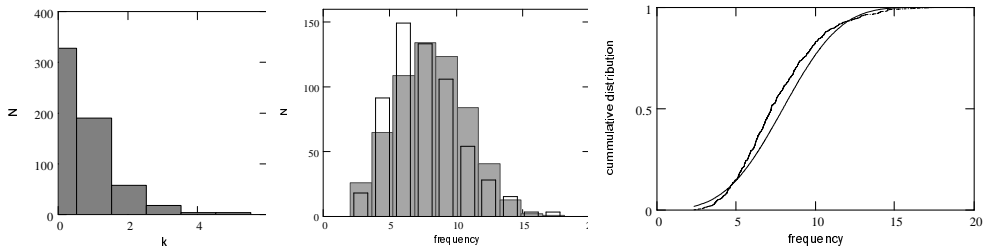

**Fig. 3.** Satisfactory restoration. Notation is same as in Fig.2.

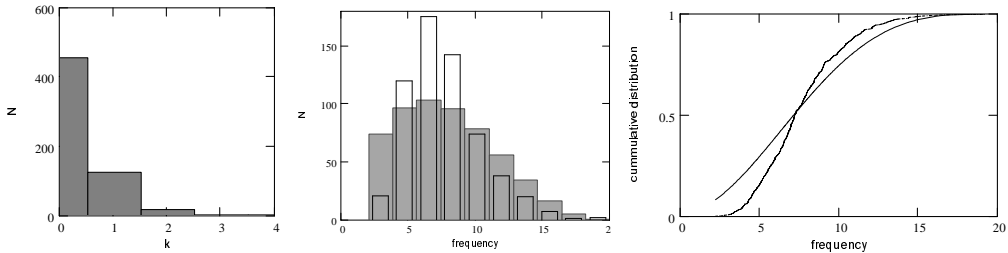

**Fig. 4.** Bad. restoration. Notation is same as in Fig.2.

The essence of the method is in the determination of statistical moments of the frequency of the SNe explosions and subsequent determination of distribution function .
2. The program of numerical simulation has been developed for testing the efficiency of the method. Numerical simulations show that even for a small number of SNe events, method allows to restore initial distribution function.